\documentclass[%
 reprint,
 amsmath,amssymb,
 aps,
]{revtex4-2}
\usepackage{xcolor}
\usepackage{hyperref}
\usepackage{graphicx}
\usepackage{dcolumn}
\usepackage{bm}

\graphicspath{{figs/}}

\begin{document}

\preprint{APS/123-QED}

\title{On morphological and functional complexity of proteinoid microspheres}


\author{Saksham Sharma}
\affiliation{Unconventional Computing Laboratory, UWE Bristol; Cambridge Centre for Physical Biology, Cambridge, UK}
\author{Adnan Mahmud}
\affiliation{Department of Chemical Engineering, Cambridge, Philippa Fawcett Drive, Cambridge CB3 0AS, UK}
\author{Giuseppe Tarabella}
\affiliation{Institute of Materials for Electronic and Magnetism, National Research Council
(IMEM-CNR), Parma (Italy)}
\author{Panagiotis Mougoyannis}
\author{Andrew Adamatzky}
\affiliation{
Unconventional Computing Laboratory, UWE Bristol, UK
}%
\date{\today}

\begin{abstract}
Proteinoids are solidified gels made from poly(amino acids) based polymers that exhibit oscillatory electrical activity. It has been proposed that proteinoids are capable of performing analog computing as their electrical activity can be converted into a series of Boolean gates. The current article focuses on decrypting the morphological and functional complexity of the ensembles of proteinoid microspheres prepared in the laboratory. We identify two different protocols (one with and one without SEM) to prepare and visualize proteinoid microspheres. To quantify the complexity of proteinoid ensembles, we measure nine complexity metrics (to name a few: average degrees $\textrm{Deg}_{av}$, maximum number of independent cycles $u$, average connections per node $\textrm{Conn}_{av}$, resistance $\textrm{res}_{\textrm{eff}}$, percolation threshold $\textrm{perc}_{\textrm{t}}$)   which shine light on the morphological, functional complexity of the proteinoids, and the information transmission that happens across the undirected graph abstraction of the proteinoid microspheres ensembles. We identify the complexity metrics that can distinguish two different protocols of preparation and also the most dense, complex, and less power consuming proteinoid network among all tested. With this work, we hope to provide a complexity toolkit for hardware designers of analog computers to design their systems with the right set of complexity ingredients guided one-to-one by the protocol chosen at the first place. On a more fundamental note, this study also sets forth the need to treat gels, microspheres, and fluidic systems as fundamentally information-theoretic in nature, rather than continuum mechanical, a perspective emerging out from recent program by Tao to treat fluids as potentially Turing-complete and thus, programmable. 
\end{abstract}

\maketitle


Proteinoids are the gels in the shape of microspheres which are the copolymers made out of amino acids of various kinds~\cite{harada1958thermal, fox1959production}. Applications of such microparticles range from producing protein-like molecules \cite{harada1958thermal}, adding fluorescent probes to the proteinoid nanoparticles (PNPs)~\cite{hadad2020engineering},  drug delivery~\cite{hadad2020engineering,haratake1998self,kolitz2018recent},  producing anti-fog additive (surbitan monooleate encapsulated into PNPs)~\cite{sason2017engineering} to using micro/nano particles in potentially delivering drug for targeted applications \cite{dhar2020self} and using as model systems for acoustomanipulation in sealed microfluidic platform \cite{} in microfluidic platform \cite{pitingolo2018beyond}. There are plentiful of protocols available in the literature to produce proteinoids, each suited to the specific application
A comprehensive list of the most widely used protocols published in the literature was introduced by us previously here \cite{sharma2022review}. There we two protocols for the production of proteinoid microspheres: (i) one based out of \textit{Fox \& Waeheldt (1968)} \cite{fox1968therml}; and (ii) another one based out of \textit{Przybylski \& Fox (1983)} \cite{przybylski1984excitable}. While the former protocol focused more upon the careful choice of the combination of several amino acids which are mixed together and heated to form proteinoids (detailed protocol in the Section \ref{sec:protocol} below), the latter protocol ensures that the proteinoid microspheres have vesicles on their periphery with controlled permeability, such that the periphery can encapsulate substances such as $\alpha-$chlorophyll which correlates to high membrane potential.

In \cite{adamatzky2021towards} we outline pathways for computing with proteinoids. We proposed that electrical spiking~\cite{przybylski1985excitable,vaughan1987thermal} of proteinoid microspheres can be use to construct neuromorphic devices~\cite{zhu2020comprehensive} and implement reservoir computing~\cite{schrauwen2007overview,lukovsevivcius2009reservoir} and mining for Boolean circuit~\cite{roberts2022mining}. In future computing devices made of proteinoid microspheres an information will need to be transferred along the microspheres' ensembles. This is why it is important to analyse the morphology of the proteinoid ensembles.

Despite plethora of work on tuning the specific details of the protocol for a desired outcome, the morphological and functional complexity of proteinoid microspheres, and the information transmission inside it, has not been studied in a greater depth till date. The most closest to this is about 50-years old study by Brooke and Fox (1971) \cite{miquel1971assembly}. In this work, the authors found that the thermal proteinoids synthesised artificially have similar morphology as the microfossils. This morphology is influenced by the presence of salts outside the proteinoids: salt water or 5\% calcium chloride induce larger size to the proteinoids than the distilled water. Acetic acid was found to induce dissolution of proteinoids: at 20\% concentration, the dissolution was initiated and at 50\% concentration, microspheres were completely dissolved. 

The morphology of the bulk of microspheres is another line of enquiry that has received less attention in the literature. Young (1965) investigated the chains which are formed by the proteinoid microspheres by inducing slight pressures on the microscopic slide, cooling the solution, altering the pH of the medium \cite{young1965morphology}. However, the morphological and functional complexity of proteinoid chains, once they are formed, was not investigated. This is because the focus of the research in 1950s was to allude the possibility that proteinoid microspheres (or coacervates) can be used as a pre-cell model to enhance our understanding of the pathways which might be responsible for the origin of cellular life. 


\section{Protocol for the preparation of proteinoids}
\label{sec:protocol}

The materials/apparatus required to prepare proteinoids \cite{panos_scirep} include: amino acids in powdered form; Tri-block heater; 3-hole round bottom flasks; nitrogen gas; magnetic stirrer; cellophane membrane for dialysis; a water bath. To start, 1.5 grams each of aspartic acid and glutamic acid is mixed and heated in the 3-hole round bottom flask at 290$^{\circ}$C initially. The temperature is step-wise increased by 10$^{\circ}$ count, until fumes start to emerge out. The exhaust in the fumehood (with N$_{2}$ gas as the inlet) is started to release the fumes out. Firstly, the powder changes from white colour to green colour, and then changes its morphology into a series of boiling microspheres. When the heating is turned off, the residue becomes solid and is extracted out and allowed to cool for half an hour. The extracted residue is put inside the Slide-A-Lyzer mini dialysis device, of 10,000 molecular weight cut-off against water as the dialysate. Dialysis is continued for 5 days until the residue consists of tiny microsphere ensembles which are visible to the naked eyes. The residue from the dialysis membrane is heated inside the vacuum oven for half an hour, to evaporate the dialysate (water) from the residue. The sample is then analysed under the transmission electron microscope. 

\section{Image analysis of proteinoid ensembles}

 \begin{figure*}[!tbp]
  \includegraphics[width=1.0\textwidth]{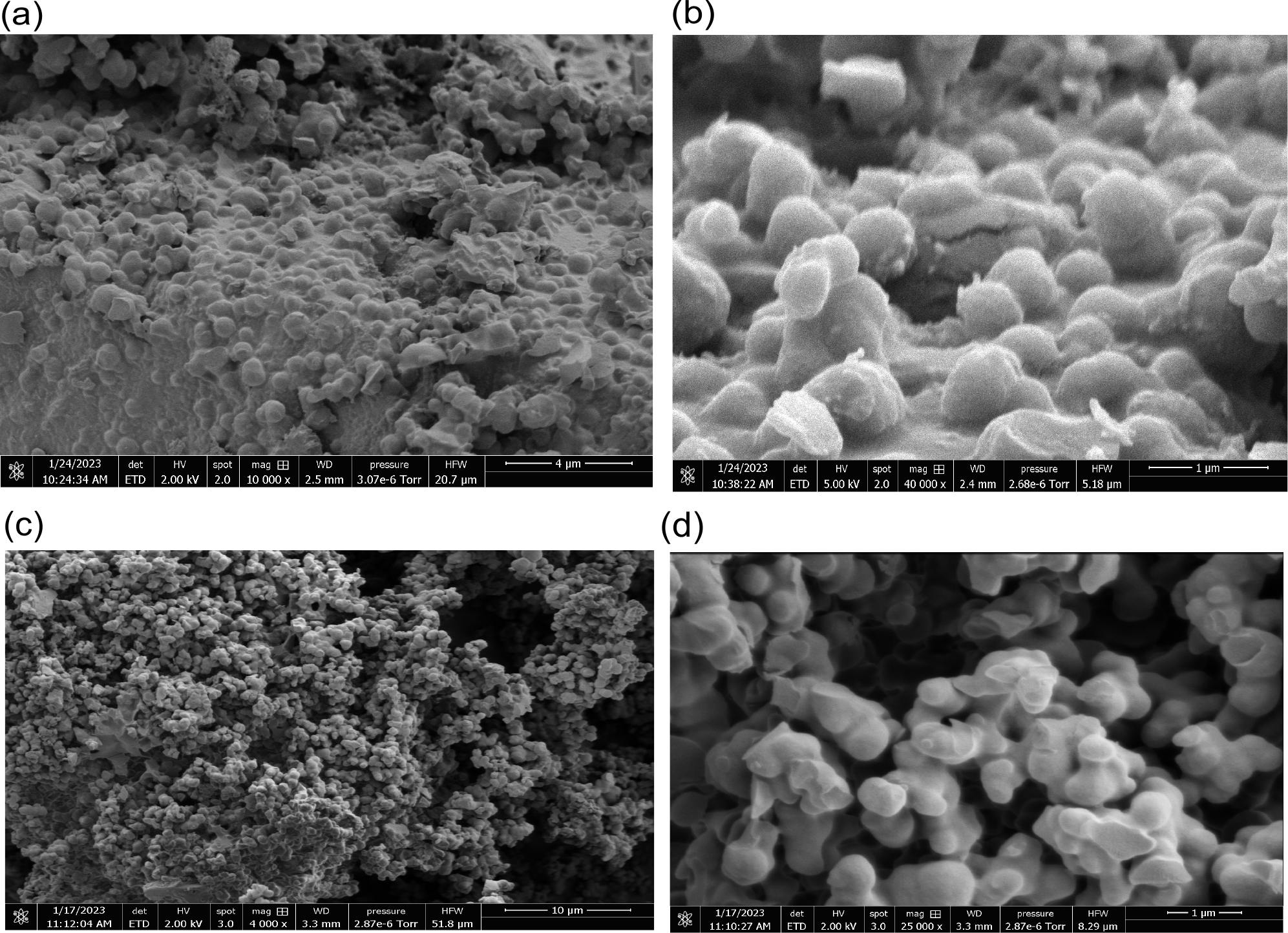}
  \caption{SEM images of the proteinoid ensembles. Equimolar mixture of L-Glutamic acid, L-asparginine, L-phenylalanine in water, at different magnification indication by the scale bars in the bottom right of each image.}
   \label{fig:fig1}
\end{figure*}

  \begin{figure*}[!tbp]
  \includegraphics[width=1.0\textwidth]{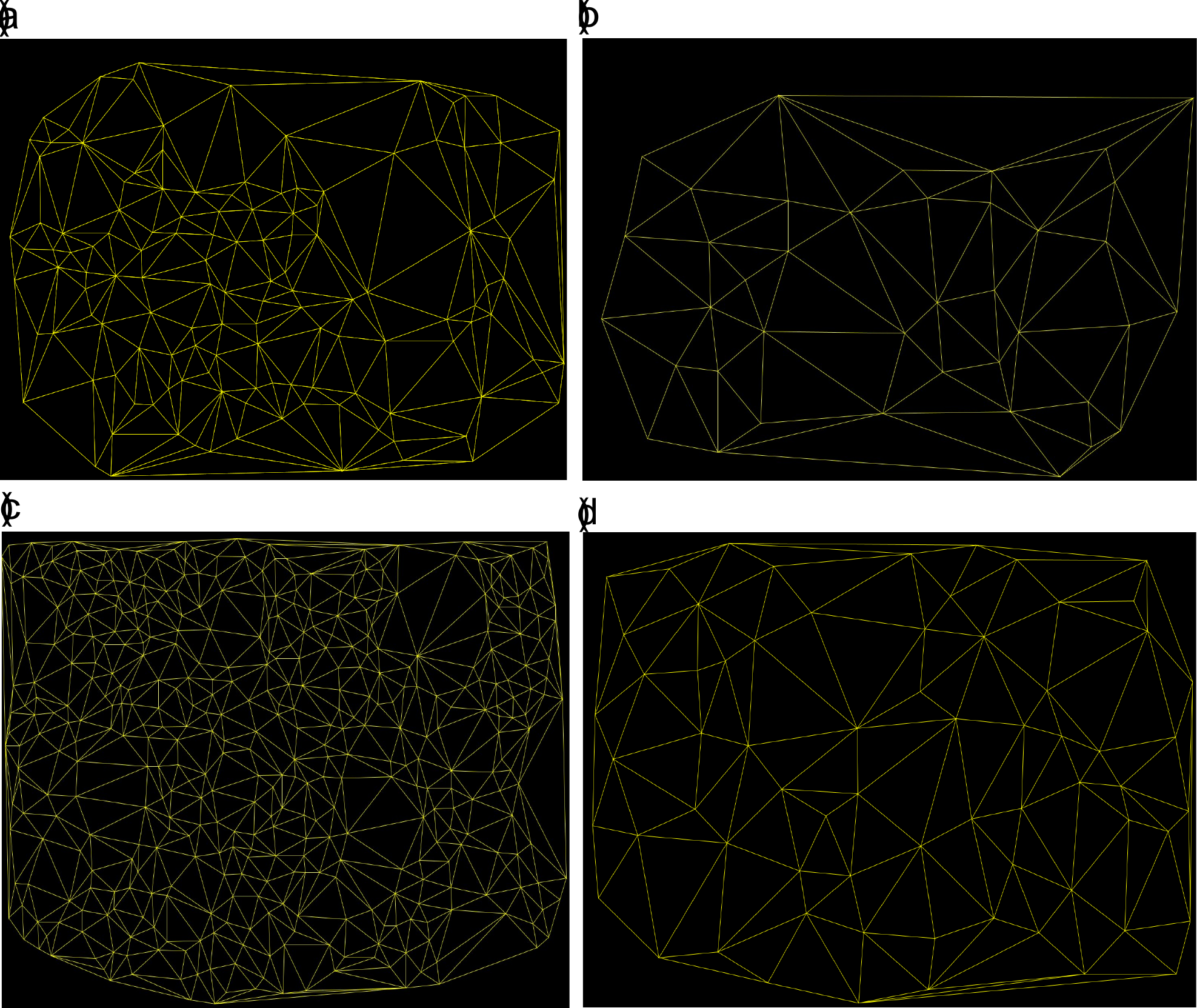}
  \caption{Undirected graphs of the images shown in Fig.~\ref{fig:fig1} generated using Relative Neighbourhood algorithm, by employing the Delaunay Voronoi plugin in open-source Fiji (ImageJ) software.}
   \label{fig:fig2}
\end{figure*}

Once the images of proteinoid microspheres are obtained using SEM (Fig.~\ref{fig:fig1}), then the next step is to analyse the granular details of the images in order to extract some relevant parameters out of it, that could enhance our physical intuition of the system. The image analysis protocol is carried out using open-source Fiji (imageJ) software, and can be divided into two major steps: 
\begin{enumerate}
    \item \textbf{Particle selection}: \newline
    On the Fiji panel, select \textcolor{blue}{Image \textgreater Adjust \textgreater Threshold}. Tick \textcolor{blue}{Dark background} and select \textcolor{blue}{Default} and \textcolor{blue}{Red} from the pull-down menus. Use the top slider so as to threshold the image until the particles are red against a dark background, then select \textcolor{blue}{Apply}. To convert this image to a binary image, select \textcolor{blue}{Process} \textgreater \textcolor{blue}{Binary} \textgreater \textcolor{blue}{Make binary}. The particle selection is done by selecting \textcolor{blue}{Process} \textgreater \textcolor{blue}{Binary} \textgreater \textcolor{blue}{Watershed}.
    \item \textbf{Relative Neighbourhood Graph sampling:}
    Once the granular images of  particles with enough clarity is obtained, the next step is to convert these images into graphs, to estimate their functional and morphological complexity. To convert raw images of particles into graphs, Delaunay triangulation (DT) need to be constructed (Fig. ~\ref{fig:fig2}). \newline
    
    The algorithm that Fiji uses to construct Delaunay triangulation is available as a plugin, accessible by selecting \textcolor{blue}{Plugins}\textgreater \textcolor{blue}{Analyze}\textgreater \textcolor{blue}{Delaunay Voronoi}. The distinctive characteristic of this algorithm is that it ensures that in the set $\mathbb{P}$ of discrete points in the graph, there exists no point in $\mathbb{P}$ which is inside the circumcircle of any triangle that is formed by the triangulation\footnote{A triangulation of a set of points $\mathbb{P}$ is a simplical complex that encompasses the convex hull of $\mathbb{P}$ and whose vertices belong to $\mathbb{P}$. } of $\mathbb{P}$.
    
    After the \textcolor{blue}{Delaunay Voronoi} option is selected, the graph that embed all the central points of the particles is constructed in the form of ROI. This graph can be saved in the \textcolor{blue}{.roi} format. It is possible to store the coordinates of the vertices and the edges of the triangulation constructed in the graph. This information will be later used to calculate graph complexity of each image, as discussed in the next section. The mean distance and the variance of points in the set $\mathbb{P}$ is also calculated using \textcolor{blue}{Delaunay Voronoi} algorithm.   
\end{enumerate}

\section{Results of the analysis}

\begin{figure*}[!tbp]
  \includegraphics[width=0.6\textwidth]{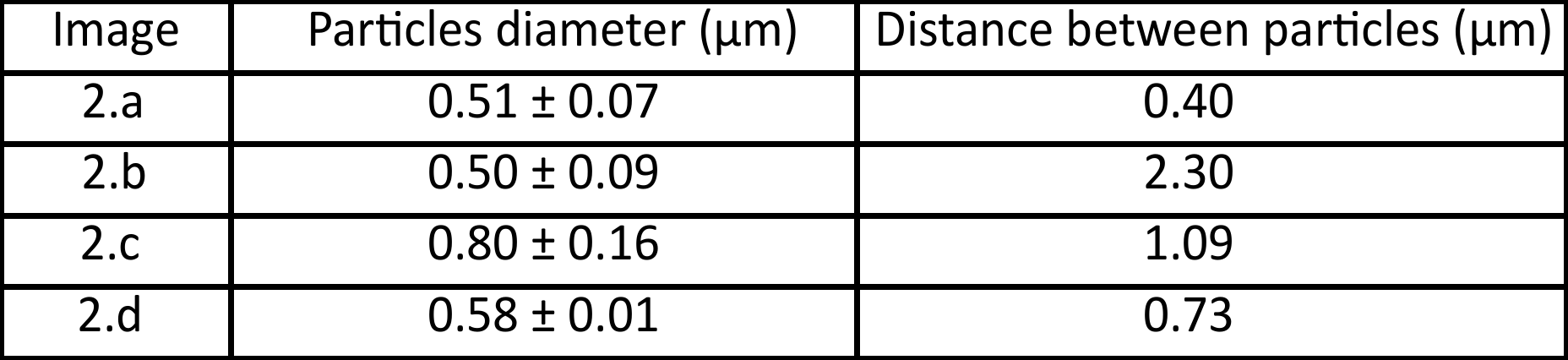}
  \caption{Results of the analysis: diameter of the proteinoid microspheres (average size and mean distance) and the distance between the particles which is recorded in the graphs.}
   \label{fig:fig3}
\end{figure*}

By employing the tools discussed in the previous section, particle sizes --- average and standard deviation --- is determined. Delaunay Voronoi triangulation method is used to determine the distances between the particles. The results are shown in Fig. \ref{fig:fig3}. Interestingly, the size recorded in the experiments is significantly lower by two orders of magnitude than the ones reported in experiments previously \cite{harada1958thermal, adamatzky2021towards}. It means the protocol for the preparation of the proteinoids \cite{panos_scirep} needs to be modified. \newline

\begin{figure*}[!tbp]
  \includegraphics[width=0.8\textwidth]{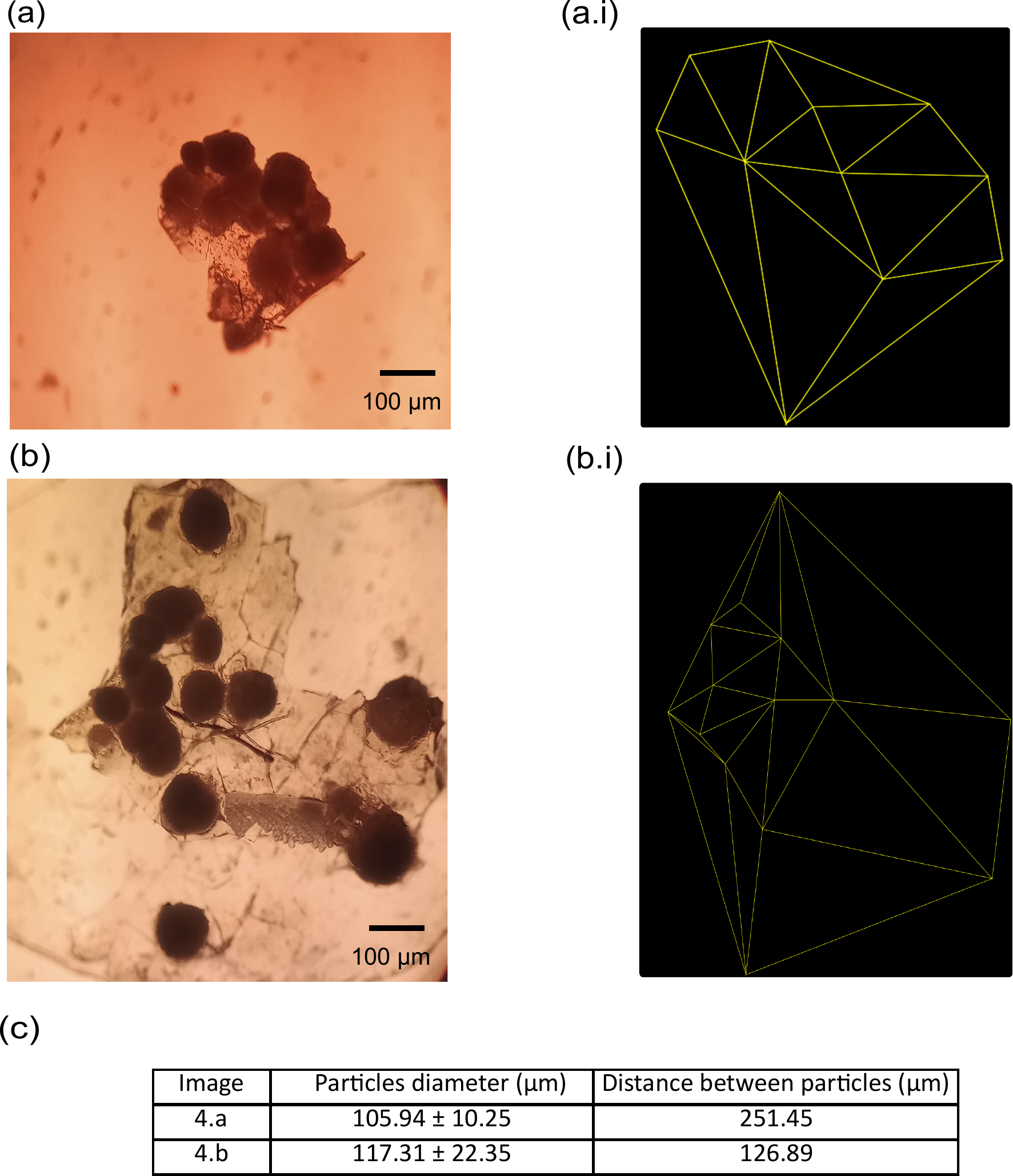}
  \caption{Proteinoid microspheres of almost 200 times larger size than the ones shown in Fig~\ref{fig:fig1}. (a,b) Proteinoids made out of 1:1 mixture of aspartic acid and L-arginine, (c) Table showing diameter of the proteinoids and distance between them recorded in the graph.}
   \label{fig:fig4}
\end{figure*}

In another batch of preparation of proteinoids (\textit{Protocol 2}) heating of the proteinoids is done at lower temperatures than before - at 70$^{\circ}$ C. Lower temperature ensures that the particles remain coarse-grained and are not destroyed (unlike at higher temperatures). Another advantage of producing proteinoids at $70^{\circ}$C is that the larger size of particles (in the order of $\sim 100$ \textmu m) allows one to use standard objective microscopes in the lab, and thus the need of using SEM vanishes. The new results for the modified protocol: images of proteinoids obtained in microscope; undirected graph corresponding to each image; results (diameter of particles and distances between them) is given in Fig. \ref{fig:fig4}. \newline

To summarise the difference in two protocols: \textit{Protocol 1} used an equimolar mixture of 
L-Glutamic acid, L-asparginine, L-phenylalanine by heating the powder (upto 290$^{\circ}$C) in the initial stage this until it gets almost black to get O(100) nanometres sized proteinoid particles. On the other hand, \textit{Protocol 2} used 1:1 mixture of aspartic acid and L-arginine to produce O(100) microns sized particles by heating the powder (upto 70$^{\circ}$C)  in the intial stage until it turns reddish brown.

\section{Complexity and information transmission analysis}

In the previous section, Delaunay triangulation method was used to determine the size of the particles and the distance between them. In this section, the discussion on the functional and morphological complexity of the proteinoid particles will be pursued. \newline

The metrics to understand the complexity of proteinoids are manifold. The primary choice behind the choice of these metrics stems from the fact that the interconnection between proteinoid particles is abstracted as undirected graphs. The primary metrics of relevance here are the total number of edges $e$ and total number of nodes (vertices) $v$ in the graph. A graph $G$ is usually characterised by the number of vertices and edges that it contains, and usually expressed as $G(e,v)$. Once the graph is defined, the secondary set of metrics comes into picture. First in the list is the \textit{average degree} given by 
\begin{equation}    \textrm{Deg}_{av}=\frac{\textrm{Total edges}}{\textrm{Total nodes}}
\end{equation}
defined and previously discussed here \cite{avgdeg} for an undirected graph. The second in the list is the maximum number of independent cycles $u$ given by 
\begin{equation}
    u=\frac{e}{2}-v+p 
\end{equation}
where $p$ is the number of subgraphs, previously discussed here \cite{indcyc} and edges are counted once (hence, divided by two). Third metric in the list is the diameter of the graph $D_{G}$ given by shortest distance between two vertices which are at the largest distance from each other, and the distance is counted in terms of nodes that connects the vertices. \newline

Apart from complexity measures, information transmission is another parameter that is important to understand the amount of information flow between nodes in a graph. An important parameter to characterise the amount of information transmission is given by the average connections per node $\textrm{Conn}_{av}$ given by 
\begin{equation}
    \textrm{Conn}_{av}= \frac{\textrm{Total connections}}{\textrm{Total nodes}}
\end{equation}
\newline
where total connections is given by the cumulative number of connections each node has in the graph. \newline

There are three more metrics that are slightly less direct to calculate that the ones mentioned above. These are, namely, total effective resistance $\textrm{res}_{\textrm{eff}}$, average shortest path $p_{short}$, and average edge length $l_{edge}$. The definitions of each of them are as follows. 

\begin{figure*}[!tbp]
  \includegraphics[width=1.0\textwidth]{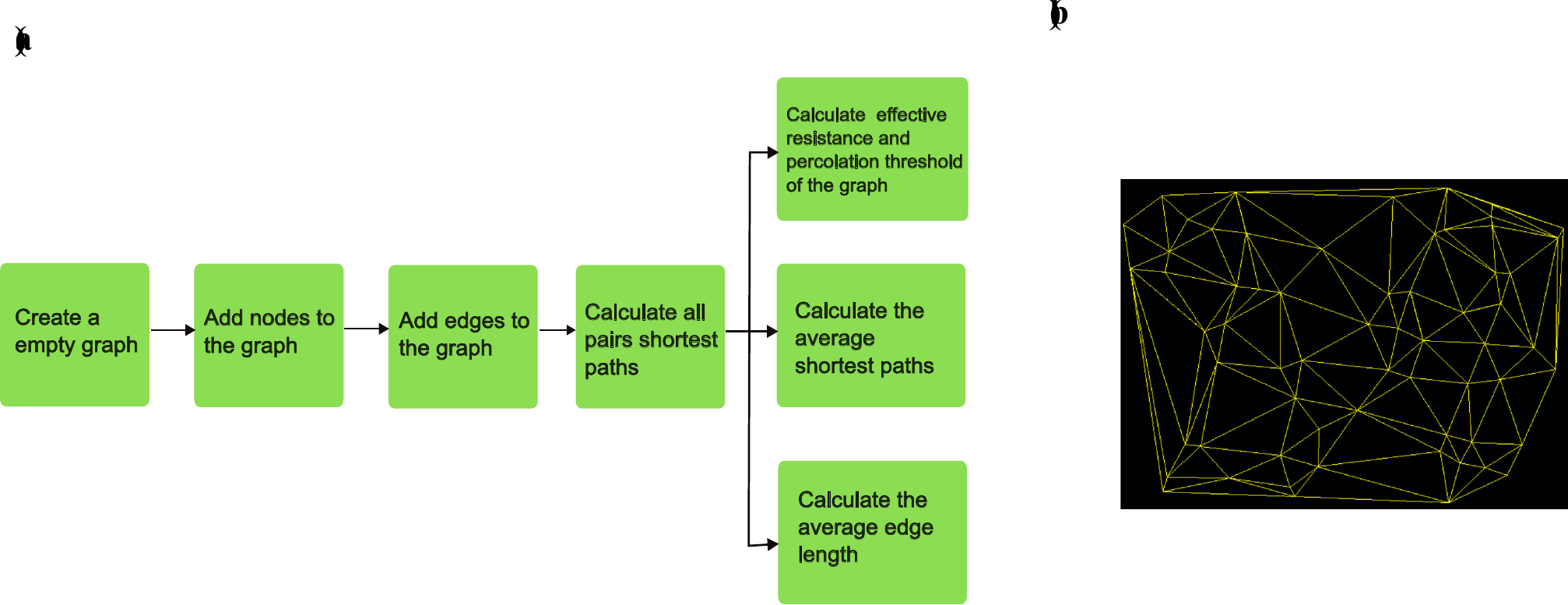}
  \caption{(a) Schematic representation of the pipeline measuring the complexity ($\textrm{res}_{\textrm{eff}}$, $p_{short}$, $l_{edge}$) of a given set of nodes, (b) Randomly generated undirected planar graph with 64 nodes.}
   \label{fig:fig5}
\end{figure*}
\begin{figure*}[!tbp]
  \caption{Table of nine complexity and information transmission metrics for the images of proteinoid ensembles captured in the present study. Image labelled `random' is shown in Fig. \ref{fig:fig5}(b).}
\includegraphics[width=0.8\textwidth]{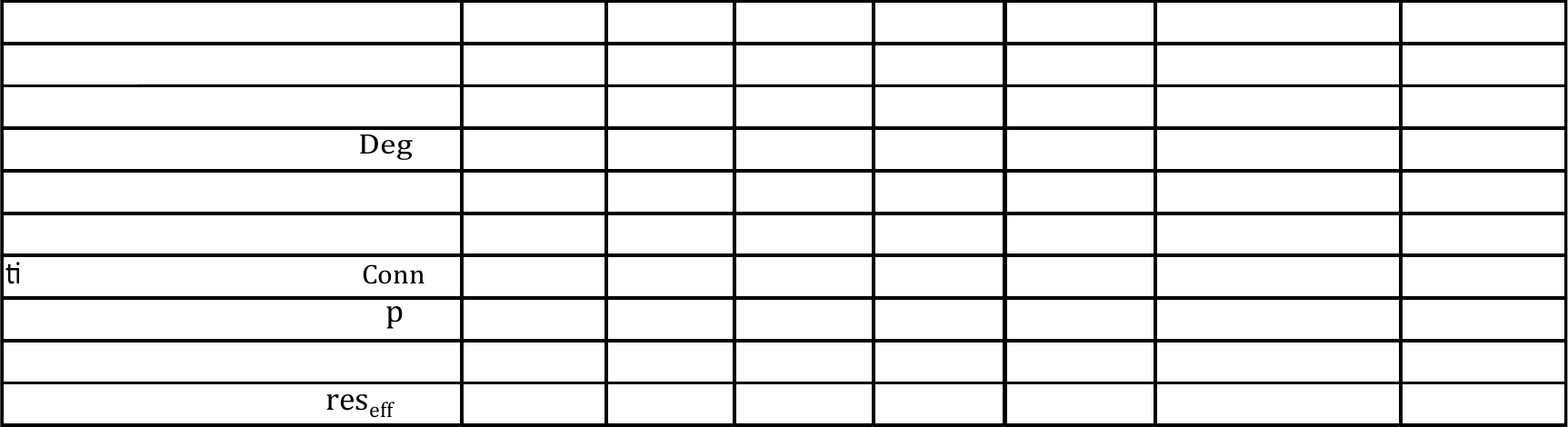}
   \label{fig:fig6}
\end{figure*}
The \textcolor{blue}{NetworkX} library in Python is used to calculate the three metrics: $\textrm{res}_{\textrm{eff}}$; $p_{short}$; $l_{edge}$ mentioned above. At the core, \textcolor{blue}{all-pairs shortest paths} is calculated using \textcolor{blue}{Floyd-Warshall algorithm}. By including intermediary nodes, the Floyd-Warshall method aims to steadily improve the shortest route estimates between all pairs of nodes \cite{floyd_warshall}. Initially, the algorithm considers the direct edge weights between nodes as the shortest path estimates. Then, it iteratively updates the estimates by considering all possible intermediate nodes. For each pair of nodes, the \textcolor{blue}{shortest path length} is extracted from the \textit{all-pairs shortest paths array}. Then the \textcolor{blue}{effective resistance} is calculated as the inverse of \textcolor{blue}{shortest path length}. The \textcolor{blue}{total effective resistance} is the sum of all \textcolor{blue}{effective resistances}. The \textcolor{blue}{total shortest paths} is calculated as the sum of all the shortest path lengths in the \textcolor{blue}{all-pairs shortest paths} array. The \textcolor{blue}{average shortest path} is then calculated by dividing the \textcolor{blue}{total shortest paths} by the number of possible pairs of nodes in the graph. The code iterates through all the edges in the graph and sums their weights (lengths) as \textcolor{blue}{total edge lengths}. The \textcolor{blue}{average edge length} is calculated by dividing the \textcolor{blue}{total edge lengths} by the number of edges in the graph. The schematic of the pipeline used to measure complexity metrics for a given set of nodes is given in Fig. \ref{fig:fig5}. Another important metric calculated in this analysis is the \textcolor{blue}{percolation threshold}. Percolation threshold $\textrm{perc}_{\textrm{t}}$ is essentially a critical value of the proportion of nodes or edges in a graph, below which the network becomes fragmented into isolated clusters, i.e., the graph transitions from being fully connected to disconnected, and above which, the graph can be seen a giant fully connected graph of roughly the same size as the system itself. The algorithm to measure is as follows:
\begin{enumerate}
    \item A set of edges are randomly removed using a variable $p$ (defined below). 
    \item For the graph, before and after, removing the edges, calculate the largest connected component (using a built-in \textcolor{blue}{nx} library function)
    \item Finally, the percolation threshold is defined as dividing the size of the largest connected component after edge removal by the size of the largest connected component before edge removal.
\end{enumerate}
\begin{figure*}[!tbp]
  \includegraphics[width=1.0\textwidth]{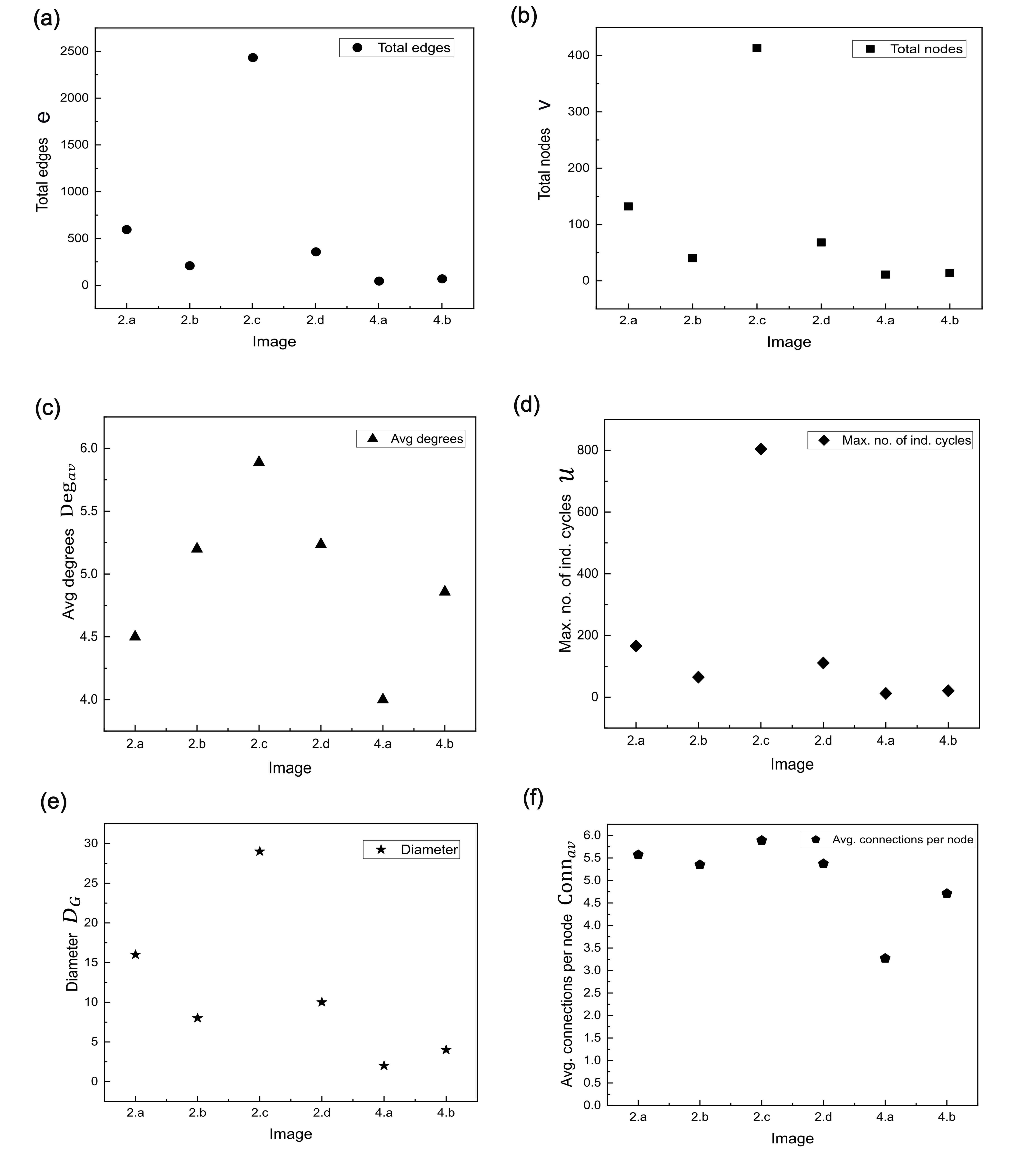}
  \caption{Plots of complexity metrics $e,v, \textrm{Deg}_{av}, p_{short}, l_{edge}, u, D_{G}, \textrm{Conn}_{av}$ against the images Fig. 2(a-d) and Fig. 4(a-b). $D_{G}$ is defined in terms of nodes.}
   \label{fig:fig7}
\end{figure*}

\begin{figure*}[!tbp]
  \includegraphics[width=1.2\textwidth]{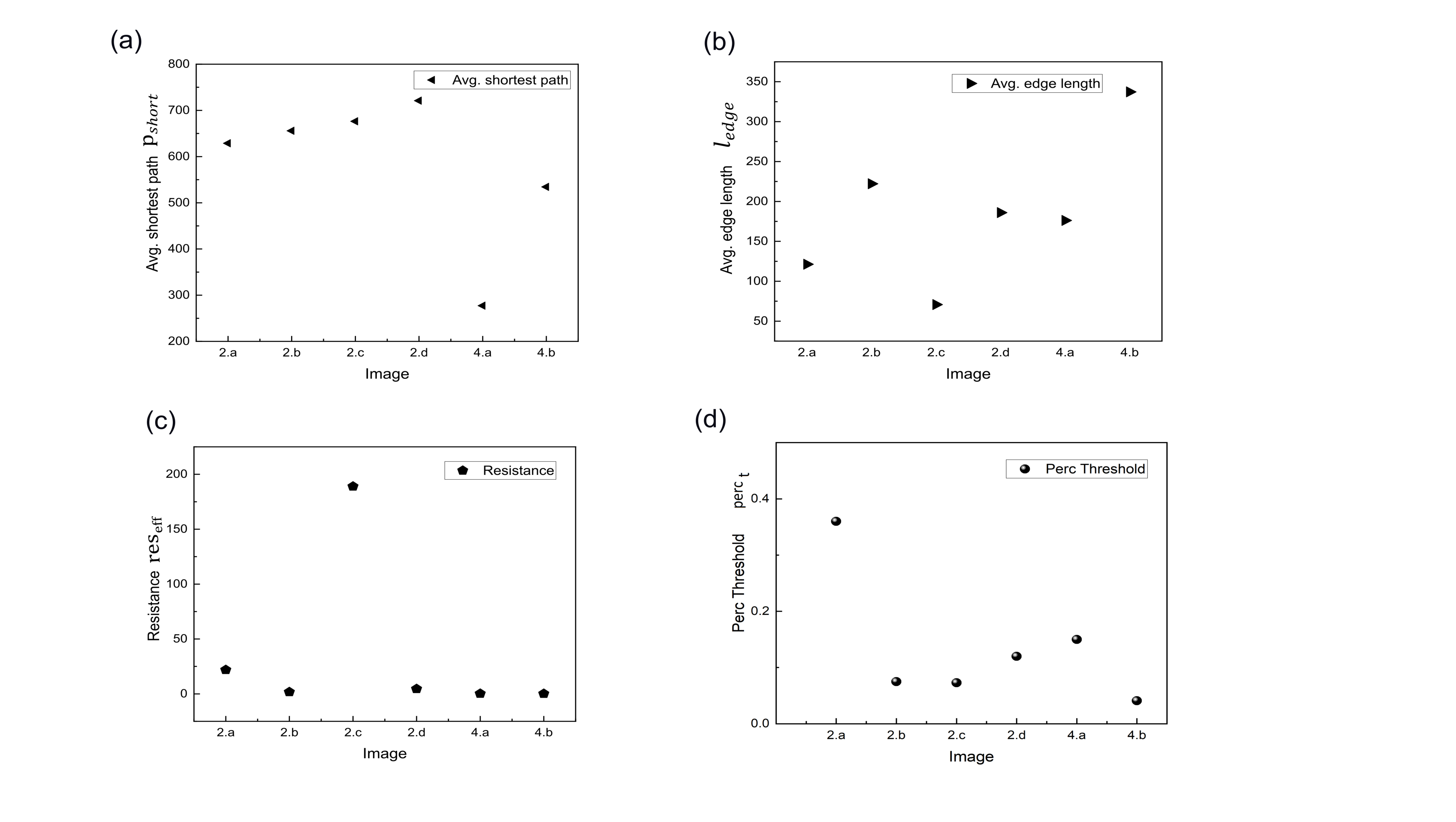}
  \caption{Plots of complexity metrics $\textrm{p}_{short},l_{\textrm{edge}}, \textrm{res}_{\textrm{eff}}, \textrm{perc}_{\textrm{t}}$ against the images Fig.~\ref{fig:fig2}(a-d) and Fig.~\ref{fig:fig4}(a-b).}
   \label{fig:fig8}
\end{figure*}

The variable $p$ represents the proportion of edges that are to be randomly removed from the graph.  To compute the percolation threshold traditionally, one needs to vary the proportion $p$ of edges that are to be removed, and then observe how the size of the largest connected component changes. One might term the percolation threshold measured, thus, as a relative percolation threshold $RPT$, a term not introduced before in the literature, which is used across all the graphs. By iteration over all the $p$ values, one \textit{observes} the breaking point.    

\section{Complexity of proteinoid ensembles: results and discussion}
Using the methods described above, the nine complexity metrics: $e,v, \textrm{Deg}_{av}, p_{short}, l_{edge}, u, D_{G},\textrm{Conn}_{av},\textrm{res}_{\textrm{eff}},\textrm{perc}_{\textrm{t}}$ are evaluated for each of the image of the proteinoid, and the results are shown in Fig.~\ref{fig:fig6}. The results for each of the parameter against the images are plotted in Fig.~\ref{fig:fig7}- \ref{fig:fig8}. \newline

From the figures, it is apparent that Fig.~\ref{fig:fig2}(c) has the highest number of total edges $e$ and total nodes $v$, making it the most dense graphs and proteinoid ensembles out of all the images tested. However, quantifying complexity using edges and nodes is not a good indicator to understand the functional complexity of a system. The only advantage of measuring $e$ and $v$ is that a set of random undirected graphs (\ref{fig:fig5}(b)) can be simulated to compare the complexity measures against, and thus, have scaled values of complexity instead of absolute. In Fig. \ref{fig:fig6}, a random undirected graoh with nodes $v=81$ and edges $e=384$ is constructed to compare against the experimental images in Fig. \ref{fig:fig2} and \ref{fig:fig4}. Highest complexity of Fig.~\ref{fig:fig2}(c) at 5.89 against the base value of 5 for random undirected graph (scaled complexity of 1.18) is also apparent through the measures of $\textrm{Deg}_{av}$; the lowest scaled complexity (average degree) is 0.8 for \ref{fig:fig4}(a). As is apparent from this comparison, the average degree metric for the \textit{Protocol 2} is significantly lower than \textit{Protocol 1}, meaning that the information flow is slower and the network is lesser robust and resilient against random failures and targeted attacks. Another important complexity metric is $\textrm{Conn}_{av}$, which is highest for Fig.~\ref{fig:fig2}(c), with a scaled complexity value of 1.08; the lowest scaled value is 0.55 for \ref{fig:fig4}(a). This again affirms that the information flow and robustness-resilience of the proteinoid hardware is highest when \textit{Protocol 1} is used as opposed to \textit{Protocol 2}. Apart from the metrics above, the diameter of the graph (in units of nodes) is highest for Fig. 2(c) (scaled value of 2.64) which means that the proteinoid ensemble in Fig.~\ref{fig:fig2}(c) is both widely and densely distributed, in comparison to the scaled value of 0.18 for \ref{fig:fig4}(a). It should be however noted that the absolute average number of connections for proteinoid ensembles fall somewhere between 4.5 and 5.89 (scaled falls in the range 0.9-1.2); in the case of Fig.~\ref{fig:fig4}(a,b) it can reduce further down to scaled 0.8. It again confirms a building hypothesis that the protocol of preparation of proteinoid has a direct influence on the range of $\textrm{Conn}_{av}$, particularly \textit{Protocol 1} builds proteinoid hardware which is more dense, widely spreaded with lower information loss and higher resilience-robustness. Apart from the such primary morphological and complexity measures, we are also interested in secondary metrics that might inform us better about the analog hardware of proteinoid ensembles. \newline

To investigate the information transmission, resistance of the graphs is shown in Fig.~\ref{fig:fig8}(c) and it is apparent that the Fig.~\ref{fig:fig2}(c) has the highest resistance: the scaled value is 42.81, thus \textit{Protocol 1} offers the maximum resistance to pass any information through itself. In comparison, Fig. \ref{fig:fig4}(b) has the lowest scaled value of resistance at 0.05, again confirming that the \textit{Protocol 2} is less preferable than \textit{Protocol 1}. In the context of information transmission, it would be safe to say that the lower resistance of proteinoid network produced using \textit{protocol 1} would mean that the energy dissipation is higher and the magnitude of electric current passing is signficantly lower by almost three orders of magnitude than the proteinoid network produced by \textit{protocol 2}. This offers a new insight into the design of analog hardware using proteinoid networks; it is not always the case that \textit{protocol 1} 
 \textit{could be} sometimes desirable for its highly complex network design. To minimise energy loss in information transmission, a less dense and more scattered proteinoid network is preferable. This insight does bolsters the importance of measuring secondary complexit metrics, instead of making conclusions only from primary metrics. Such secondary measures of complexity metrics are useful for our analog computing studies using proteinoid microspheres in the future, where we would be interested in converting the information transmission occurring through the proteinoid microspheres to a series of Boolean gates, and these logical gates would become a hidden, underlying information-theoretic \textit{language} of the microspheres. \newline

Since Fig.~\ref{fig:fig2}(c) is the most complex and dense image with lower rate of information transmission, it has the least edge length as shown in Fig.~\ref{fig:fig8}(b). It should be however noted that the average shortest path is not a reliable measure of complexity of the graph within the images which are prepared using the same protocol (Fig.~\ref{fig:fig2}(a-d)), however, it still can be used to compare different protocols of preparation: $p_{short}$ for Fig.~\ref{fig:fig4}(a-b) (\textit{protocol 2}) is significantly lower than Fig.~\ref{fig:fig4}(a-b) (\textit{protocol 1}). It is to be noted that the average shortest path and average edge length complexity metrics do not offer a direct insight into the information transmission inside the proteinoid networks. However, they do help a hardware designer to make sure that the proteinoid networks, thus designed, are consistent over multiple factory makes. It is good to again reinstate that such metrics are useful only to ensure the designer to produce identical proteinoid networks as final customer products such that analog computing operation becomes device agnostic. This feature is important to consider while designing industry production pipeline for analog hardware and computing systems made out of proteinoid networks. \newline

The last important complexity metric to consider is the percolation threshold $\textrm{perc}_{t}$. The scaled value of $\textrm{perc}_{t}$ is lowest (0.05) for \textit{Protocol 2} and highest (0.41) for \textit{Protocol 1}. This means that it is relatively easier for the proteinoid network produced using \textit{Protocol 2} to become a single fully connected piece of hardware carrying out similar kind of operations. Networks produced using \textit{Protocol 1}, on the other hand, have more tendency to remain as isolated clusters with as little connection as possible. The usefulness of having isolated clusters is that the proteinoid hardware can be useful to perform analog computing processes that require parallel operations (read, \textit{cores}, in digital context) to expedite the computing task. It also allows to reduce the chances of failure in completion of the task if some clusters fail to perform well - the system, on the whole, can be designed to not effectively fail, and still complete the task. 

\section{Conclusion}
This study identifies key complexity metrics that can characterise the computational and design complexity for the functioning of proteinoid networks as potential chemical hardware that can aid in performing analog computing tasks. Analog computing, albeit initially popular in early 20th century, is again coming back into popularity in early 21st century, because of: (i) rise in ML algorithms that do not need to be necessarily accurate 100\% in their output, and (ii) Moore's law about to come to an end, hence, a demand for non-digital computing hardwares. Proteinoids, which are essentially gels, made from poly(amino acids) are shown to be potential contender in this direction. \newline

We devised two different protocols to produced proteinoids; \textit{Protocol 1} to develop O(100) micron sized and \textit{Protocol 2} to develop O(100) nanometer sized particles. We used primary and secondary complexity metrics to investigate the functional complexity of such microspheres, where the function of concern is the information transmission, had the proteinoid networks are being subjected to performing arbitrary analog computing task. \textit{Protocol 1} performed better when it comes down to characterising system in terms of average degrees, average connections per node, and percolation threshold. However, \textit{Protocol 2}, despite being being less dense and less complex, is more useful to aid in easy flow of information across the system with as little power dissipation as possible. It would be, thus, safe to conclude that the protocol that would meet the industry requirement would fall somewhere between \textit{Procotol 1} and \textit{Protocol 2}, such that the complexity metrics measured above, especially the average resistance and the percolation threshold, are higher and consistent in the final product.

\section{Future directions and perspective}

After the morphological and functional complexity of the proteinoid microspheres is established and convenient protocol for the preparation of consistent and efficient analog hardware is established, we are planning to venture into experiments (using voltage-sensitive dyes) on mapping binary sequences and QR codes, represented as spikes of electrical activity, by ensembles of proteinoid microspheres. This would enable industry specialist to treat such chemical systems and their inherent language of electrical signals as no less than a potential hardware that can carry information across through its network and build fundamental block of analog computing - in the form of logic gates and QR codes - thus enabling them to customise soft matter fluidic systems in performing analog computing and machine
learning tasks. \newline

The broader scope of the work is in aiming towards relinquishing a mathematical treatment of microdroplets, because of the reasons that the mathematical continuum model of treating \textit{arbitrary}\footnote{It is to important to note that analytical solution of Navier-Stokes equations are possible for specific, special, and well-chosen cases. When it comes down to arbitrary choice of problem, \textit{i.e.}, arbitrary initial conditions of velocity fields and arbitrary domain, then, in such cases, completely analytic and general $C^{\infty}$ solution do not exist. Hence, the potential departure to information-theoretic solution, that perhaps is divisible into series of logic gates, of Navier-Stokes equations is sought after.} system of fluids or gels is inherently flawed as argued recently \cite{sharma2022complexity}. This is line with a recent result by a group of mathematicians that the Euler equations are Turing-complete, and thus fluids should be seen as fundamentally programmable instead of analytically tractable. The reason why analytical tractable of arbitrary fluidic or gel physical systems is not possible is because the continuum equations do not
take into account the atomic/molecular contributions within the system (argued here \cite{sharma2022navier}) and in cases where such contributions are not taken, physical manifestation of mathematical singularities plays out. This analytical intractability has been the key reason why the millennium problem of the 21st century - Navier-Stokes regularity problem \cite{fefferman2000existence} - still remain an elusive and unsolved problem among the community of functional analysts. It can be safely conjectured that a strong combined effort by theoretical computer scientists and logicians, apart from mathematical physicists, would be needed to solve the Navier-Stokes millennium problem. With this theme in mind, the current study also plans to investigate the design of logical gates in our upcoming work that can represent information transmission in proteinoid microspheres. Instead of using continuum mechanics to represent information in this system, we would be mindful to use
a well-designed system of Boolean gates to do the same, as suggested almost a decade ago by Tao’s fluid program \cite{tao2016finite}, inspired from Conway’s Cellular Automaton \cite{adamatzky2010game}.

\section*{Code}

The github code used in the article to estimate $p_{short}$, $l_{edge}$, $\textrm{res}_{\textrm{eff}}$ is available \href{https://github.com/Adnan1729/complexity_analysis}{\textcolor{red}{here}} and is written solely by A.M.

\bibliography{proteinoidbib1}

\end{document}